\documentclass{article}

\usepackage{PRIMEarxiv}

\usepackage[utf8]{inputenc} 
\usepackage[T1]{fontenc}    
\usepackage{hyperref}       
\usepackage{url}            
\usepackage{booktabs}       
\usepackage{nicefrac}       
\usepackage{microtype}      
\usepackage{lipsum}
\usepackage{fancyhdr}       
\usepackage{graphicx}       
\graphicspath{{media/}}     

\usepackage{multirow}%
\usepackage{amsmath,amssymb,amsfonts}%
\usepackage{amsthm}%
\usepackage{mathrsfs}%
\usepackage[title]{appendix}%
\usepackage{xcolor}%
\usepackage{textcomp}%
\usepackage{manyfoot}%
\usepackage{booktabs}%
\usepackage{algorithm}%
\usepackage{algorithmicx}%
\usepackage{algpseudocode}%
\usepackage{listings}%
\usepackage[numbers]{natbib}
\usepackage{chemformula} 
\usepackage{bigfoot}
\usepackage{commath}
\usepackage[markup=underlined]{changes}
\definecolor{myred}{RGB}{255, 0, 0} 
\setaddedmarkup{\textcolor{myred}{#1}} 

\newcommand{\specialhline}[1]{\noalign{\global\arrayrulewidth#1}\hline
	\noalign{\global\arrayrulewidth.4pt}}

\title{Dielectric Tensor Prediction for Inorganic Materials Using Latent Information from Preferred Potential}

\author{
  Zetian Mao\thanks{Work done during the internship at Preferred Networks, Inc.}\\
  Graduate School of Frontier Sciences \\
  The University of Tokyo \\
  5-1-5 Kashiwanoha, Kashiwa, 277-8561, Chiba, Japan\\
   \And
  WenWen Li\footnote{Corresponding author.}, Jethro Tan \\
  Preferred Networks, Inc. \\
  1-6-1 Otemachi, Chiyoda-ku, 100-0004, Tokyo, Japan \\
  \texttt{wenwenli@preferred.jp} \\
}

\begin{document}
\maketitle

\begin{abstract}
Dielectrics are crucial for technologies like flash memory, CPUs, photovoltaics, and capacitors,
but public data on these materials are scarce, restricting research and development.
Existing machine learning models have focused on predicting scalar polycrystalline dielectric constants,
neglecting the directional nature of dielectric tensors essential for material design.
This study leverages multi-rank equivariant structural embeddings from a universal neural network potential to enhance predictions of dielectric tensors.
We develop an equivariant readout decoder to predict total, electronic, and ionic dielectric tensors while preserving O(3) equivariance, 
and benchmark its performance against state-of-the-art algorithms.
Virtual screening of thermodynamically stable materials from Materials Project for two discovery tasks,
high-dielectric and highly anisotropic materials, identifies promising candidates including Cs\textsubscript{2}Ti(WO\textsubscript{4})\textsubscript{3} (band gap $E_g=2.93 \mathrm{eV}$, dielectric constant $\varepsilon=180.90$) and CsZrCuSe\textsubscript{3} (anisotropic ratio $\alpha_r = 121.89$). 
The results demonstrate our model's accuracy in predicting dielectric tensors and its potential for discovering novel dielectric materials.
\end{abstract}


\section{Introduction}

Dielectrics are an essential group of materials identified by their distinct electronic characteristics. 
They find extensive applications in various fields such as photovoltaics~\cite{brebels2017high}, energy storage~\cite{wu2022advanced,zha2021polymer}, 
microwave communication~\cite{hill2021perspective, shehbaz2023recent}, etc.,
in contemporary society. 
When subjected to an external electric field,
a dielectric material is polarized by internally generating electric dipole moments.
The dielectric constant $\varepsilon$, also known as permittivity, 
is a measure of such dielectric effect described by the proportionality 
between the externally applied electric field to the generated field within 
the material:
\begin{equation}
	E_i^{\textrm{int}}=\sum_{j}\varepsilon_{ij}^{-1}E_{j}^{\textrm{ext}}
\end{equation}
where $E^{\textrm{int}}$ and $E^{\textrm{ext}}$ are the internal and external electric, respectively,
and the indices $i,j\in\{1, 2, 3\}$ refer to the directions in the 3D space.
The $3\times3$ dielectric tensor has a minimum of 1 and a maximum of 6 independent
elements for various types of systems due to the crystal symmetry.
It can be represented by a sum of two components: the electronic contribution ($\varepsilon^{\infty}$) and ionic contribution ($\varepsilon^0$),
which are generated by electron cloud distortion and atomic displacement, respectively, i.e.,
$\varepsilon_{ij} = \varepsilon^{\infty}_{ij} + \varepsilon^0_{ij}$.
Both high and low dielectric constants play an essential role in 
material design. 
For instance, high-$\kappa$ dielectrics enable higher-density charge storage to shrink device sizes, 
while low-$\kappa$ dielectrics are important to device packaging by minimizing signal interference 
between neighboring interconnects.

As of the present, only a limited number of materials in few hundreds have had their dielectric constants measured~\cite{petousis2017high}
because of the requirement of huge experimental efforts.
In the past decades, first-principle calculations played a critical role for 
accelerating the discovery of new materials~\cite{hautier2012computer}.
Density functional theory (DFT)~\cite{hohenberg1964inhomogeneous,kohn1965self} 
has gained significant popularity due to its commendable balance between computational speed and accuracy for simulating materials in silico. 
This has, in turn, facilitated the emergence of a wide range of diverse new materials~\cite{choudhary2020high,greeley2006computational}.
Similarly, density functional perturbation theory (DFPT) serves as a simulation tool 
for establishing the relationship between structure and dielectric properties,
and the calculated data in the order of thousands are publicly accessible for material exploration~\cite{petousis2016benchmarking,petousis2017high}.
However, the DFT-based methods entail significant computational cost,
making them less feasible for high-throughput screening,
and it is only applicable to small systems with the number of atoms typically less than 1000~\cite{chen2020critical}, 
since its complexity scales poorly with the number of particles. 
Alternatively, machine learning is becoming an important research tool,
providing an alternative method for quantitative structure property relationship (QSPR) analysis for material science,
There are numerous successful works in designing materials for various applications,
including fluorescent molecules~\cite{sumita2022novo}, electrets~\cite{mao2023ai,zhang2021discovery},
DDR1 kinase prohibitors~\cite{zhavoronkov2019deep}, thermal-electric materials~\cite{oliynyk2018discovery}, etc.

Several existing works made efforts to make predictions for dielectric property directly from 
crystal/molecule descriptors by using the machine learning model.
Umeda et al. adopted a random forest (RF) model to estimate experimental 
dielectric constants with an error range as large as 50\% in the logarithmic scale
on the purpose of developing ferroelectric material~\cite{umeda2019prediction}. 
Takahashi et al. studied the prediction for static dielectric constants of 
metal oxides and analyzed the importance of descriptors using RF models.
Lin et al. employed a simple Gradient boosting regression (GBR) model
to predict the electronic contribution of polycrystalline dielectric constants with RMSE of 0.824 and did analysis 
on feature importance for $\textrm{ABO}_3$-type compounds~\cite{lin2021accelerated}. 
Similarly, Kim et al. used a GBR model to predict both the electronic and ionic contribution of $\textrm{ABO}_3$-type perovskites, 
achieving RMSE of 0.12 and 0.26 respectively~\cite{kim2022prediction}.
Although these works proved the strong ability of machine learning models to rapidly estimate dielectric properties on unseen materials 
and are undoubtedly helpful to guide material designers before conducting wet experiments,
there are still several limitations that hinder their full potential for exploration in the dielectric material space.
These tasks focus on the average polycrystalline dielectric constants approximated from the dielectric tensor, given by
\begin{equation}
	\varepsilon=\sum_{i=1}^3\lambda_i/3
	\label{eq:scalar dielectric}
\end{equation}
where $\lambda_i$ is the eigenvalue of $\boldsymbol{\varepsilon}$.
The Matbench dataset~\cite{dunn2020benchmarking} includes the dielectric task for benchmarking various material property models. 
The property in this task is the refractive index ($\eta$) which is related to the electronic dielectric constant, as given by $\eta=\sqrt{\varepsilon^{\infty}}$.
Various GNN-based models are evaluated on this task.
However, these works ignored the inherent anisotropy in crystal materials. Additionally, the off-diagonal elements in the dielectric tensor may reach large values and sometimes cannot be neglected in practical~\cite{sturm2015dielectric}, 
thus prediction of the dielectric constants as tensors assists human researchers to understand the behavior of materials in the presence of a directed electric field.
To distinguish, dielectric constants as scalars and as tensors are denoted with $\varepsilon$ and $\boldsymbol{\varepsilon}$, respectively.
$\boldsymbol{\varepsilon}$ could provide crucial comprehensive measures for the designs of electronic devices in various applications.

In recent years, graph neural networks (GNN) has been becoming a powerful tool 
for rapid atomic simulation in material science~\cite{battaglia2018relational,gilmer2017neural}
as graphs are a natural representation for crystals and molecules 
where nodes are atoms and edges are bonds between atom pairs.
Many great GNN-based models~\cite{chen2019graph,schutt2018schnet,gasteiger2020directional,liao2022equiformer,batzner20223,schutt2021equivariant} 
are proposed and demonstrated their efficacy in accurately establishing a quantitative structure-property relationship.
Recent works proposed the application of equivariant graph neural networks (EGNN) in materials science~\cite{takamoto2022teanet,schutt2021equivariant}
and discussed the necessity of equivariant representations to enhance model expressivity.
\textit{TeaNet}~\cite{takamoto2022teanet} is an EGNN architecture that satisfies 
special Euclidean group E(3) equivariance including translation, rotation and reflection.
Based on \textit{TeaNet} with certain modifications,
PreFerred Potential (\textit{PFP})~\cite{takamoto2022towards} is a universal neural network potential (NNP) model trained by massive data and has been released for multiple versions.
What especially makes \textit{PFP} robust for universal interatomic interactions
is its enhanced representations of the local electronic environments centered around atoms and bonds
by converting raw input data (atomic numbers and positions) to rank-0,1,2 tensors (i.e., scalars, vectors and tensors).
Pretrained on the original dataset with 22 million diverse structures (calculated by 1144 GPU years),
the up-to-date \textit{PFP} covers 72 elements in the periodic table for
empirical universal interatomic potential simulation with strong transferability~\cite{takamoto2023towards}.
We refer readers to~\cite{takamoto2022towards} for more details about this original training dataset across all systems including molecular, crystal, slab, cluster, adsorption, and disordered structures.
Its abundant intermediate interatomic information in pretrained layers of graph convolutional networks (GCN) 
and inclusion of equivariant rank-2 tensor features make it 
potentially suitable to transfer to diverse tensorial regression tasks with relatively limited datasets such as dielectric tensors.
During the same period as this study, Lou et al.~\cite{lou2024discovery} proposed using equivariant graph neural networks to learn irreducible representations for predicting dielectric materials with highly anisotropic electronic dielectric tensors. 
In contrast, our study leverages transfer learning of pretrained geometric features across properties with varying tensor orders to enable dielectric tensor predictions over a broader numerical range.
Moreover, instead of focusing exclusively on electronic components, we evaluate our models on electronic, ionic, and total dielectric tasks, demonstrating state-of-the-art performance in predicting both scalar and tensor properties.

In this study, we propose \textit{Dielectric Tensor Neural Network}(\textit{DTNet}) to predict the three types of dielectric tensors,
$\varepsilon_{ij}^\infty$, $\varepsilon_{ij}^0$ and $\varepsilon_{ij}$, 
for inorganic materials across 72 supported elements with the equivairance to the input structure.
By leveraging a pretrained PFP model as an efficient and expressive encoder,
we demonstrated that the pretrained \textit{PFP} processes a rich set of 
high-order latent compositional and structural information for tensorial property prediction.
The ablation study shows that both latent atomic features and bond features from \textit{PFP} contribute to the final prediction.
To evaluate the performance of \textit{DTNet}, 
we compared it against existing methods, PaiNN~\cite{schutt2021equivariant}, M3GNet~\cite{chen2022universal} and MatTen~\cite{wen2023universal}, on the dataset obtained from Materials Project (MP)~\cite{jain2013commentary}.
Finally, we applied our model to virtual screening for two discovery tasks: identifying high-dielectric materials and highly anisotropic dielectrics.

\section{Results}

\subsection{Overview}

\begin{figure}[tb]
	\centering
	\includegraphics[width=\textwidth]{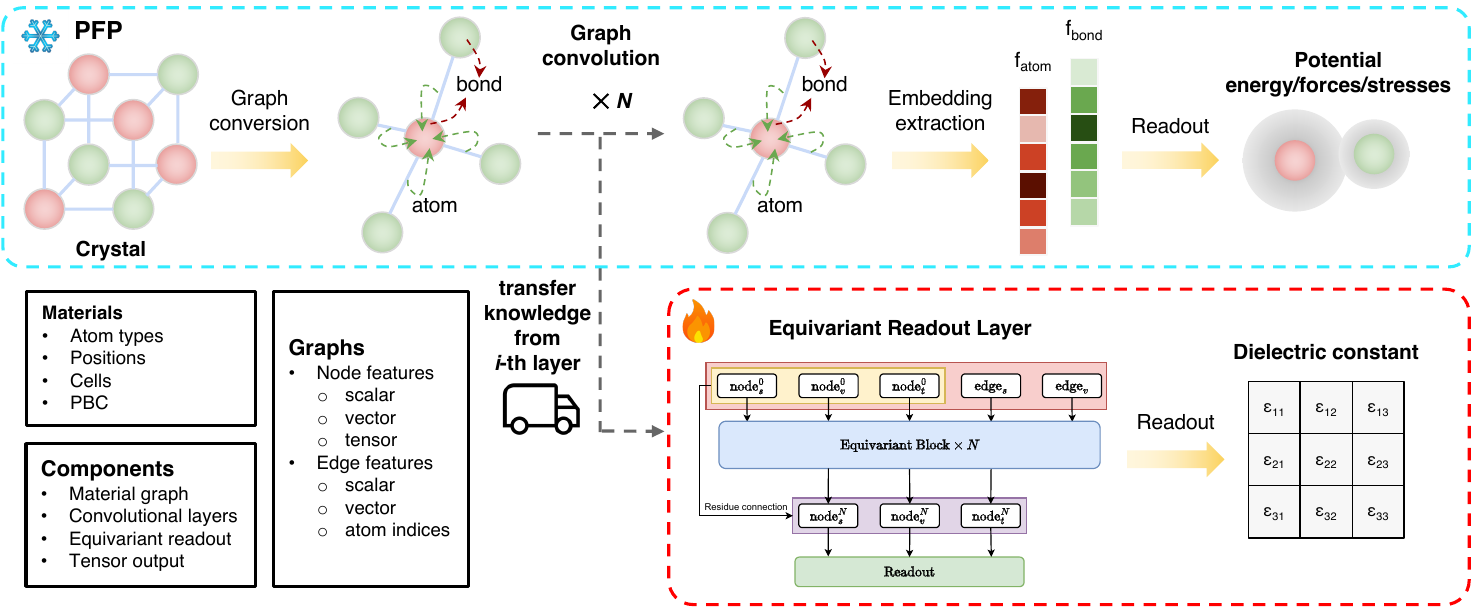}
	\caption{Diagram of the equivariant model for dielectric tensor prediction. The input graph is represented by initialized atom attributes $\mathcal{V}=\{(\boldsymbol{a}_s, \boldsymbol{a}_v, \boldsymbol{a}_t)_i\}_{i=1:N^a}^0$ and bond attributes $\mathcal{E}=\{(\boldsymbol{b}_{s,{\{i,j\}}}, \boldsymbol{b}_{v,{\{i,j\}}})_k\}_{k=1:N^b}^0$ between atoms with indices of $i$ and $j$, where $N^a$ and $N^b$ are number of atoms and bonds. The intermediate knowledge output $\{(\boldsymbol{a}_s, \boldsymbol{a}_v, \boldsymbol{a}_t)_i\}_{i=1:N^a}^n$ and $\{(\boldsymbol{b}_{s,{\{i,j\}}}, \boldsymbol{b}_{v,{\{i,j\}}})_k\}_{k=1:N^b}^n$ from the $n$-th layer GCN of \textit{PFP} is fed into an equivariant readout block to perform message interactions among different rank representations for a downstream task, i.e., dielectric tensor prediction here.}
	\label{fig:overview}
\end{figure}

\begin{figure}[tb]
	\centering
	\includegraphics[width=\textwidth]{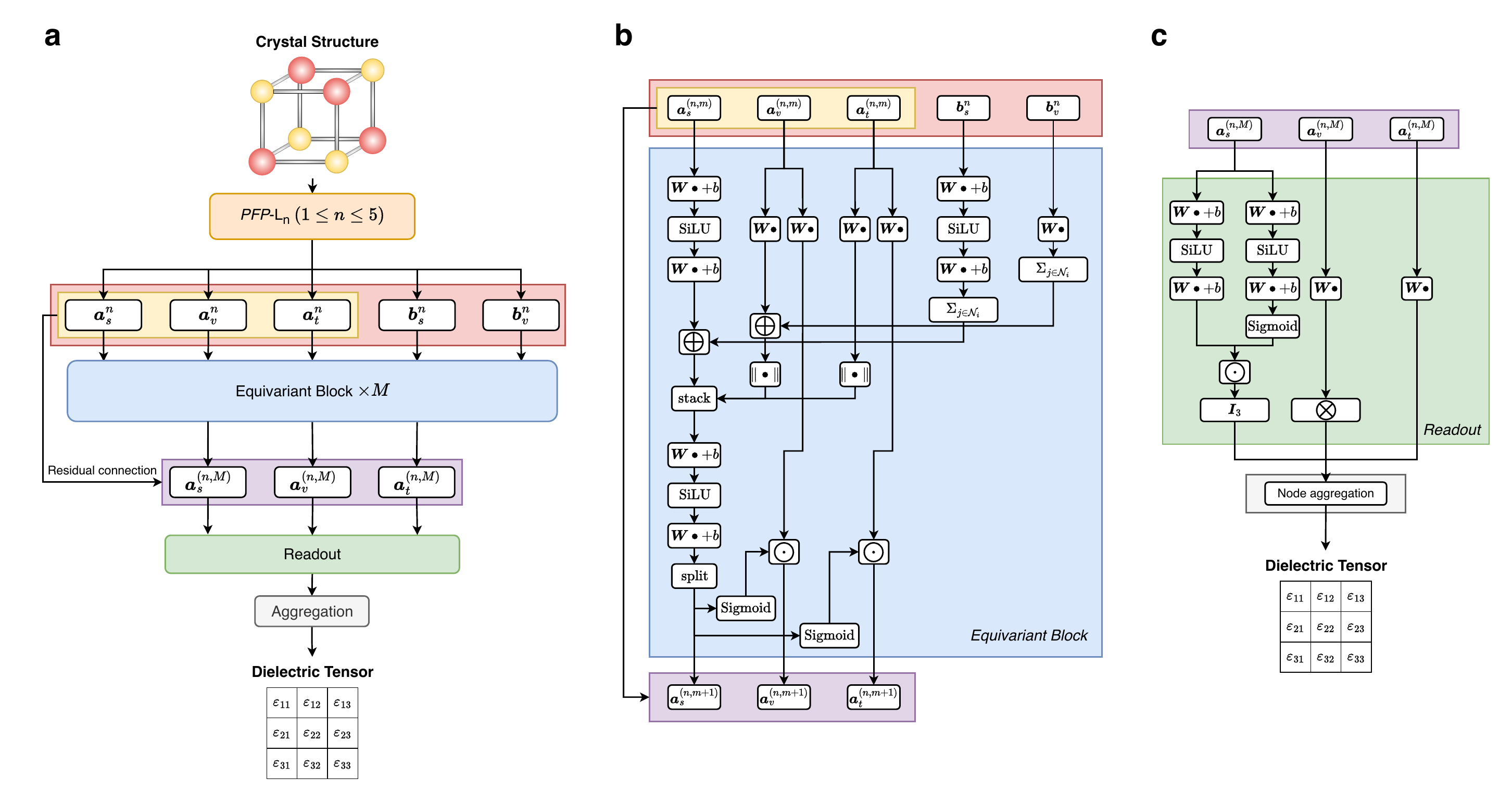}
	\caption{Schematic architecture of the equivariant dielectric model \textit{DTNet}. (a) Overview of the \textit{DTNet} framework; (b) Implementation of an equivariant block in \textit{DTNet}; (c) Implementation of the readout block for dielectric tensor output}
	\label{fig:model}
\end{figure}

The challenge of GNN in materials science is that they usually require a large number of data to prevent models from overfitting,
while such chemical data are generally expensive to collect either from simulation or experiments.
Additionally, the numbers of calculated data for diverse properties can have large difference.
For example, by the time of this work, there are more than $\sim$154k DFT-relaxed material structures with energies in MP, but only $\sim$7k (<5\%) of them are available with dielectric constant data.
In this work, we took a \textit{PFP} as the parent encoder model, i.e., $N=5$.
And for convenience, we denote the \textit{PFP} with the $n(1\leq n\leq5)$-th GCN layer as featurizer as \textit{PFP}-L\textsubscript{1}, ...,\textit{PFP}-L\textsubscript{5}.
Our method leverages a \textit{PFP} model consisting of 5 GCN layers, 
which was pre-trained on 22 million first-principle calculations of potential energies, 
as a multi-order feature encoder to generate universal compositional and structural information for the second-order dielectric property prediction (see Figure~\ref{fig:overview}).
We design a separate model (see Fig.\ref{fig:model}) to capture interactions among the intermediate knowledge from \textit{PFP}-$\mathrm{L}_n$
and make predictions on downstream tasks without the need to finetune parameters in the parent \textit{PFP} model.
This design ensures the effective utilization of high-order common structural and compositional information, leading to improved performance even with a relatively small region of available training data.
Details about the module implementation are available in the Methods Section.

\subsection{Equivariance}

\begin{figure}[tb]
	\centering
	\includegraphics[width=0.4\textwidth]{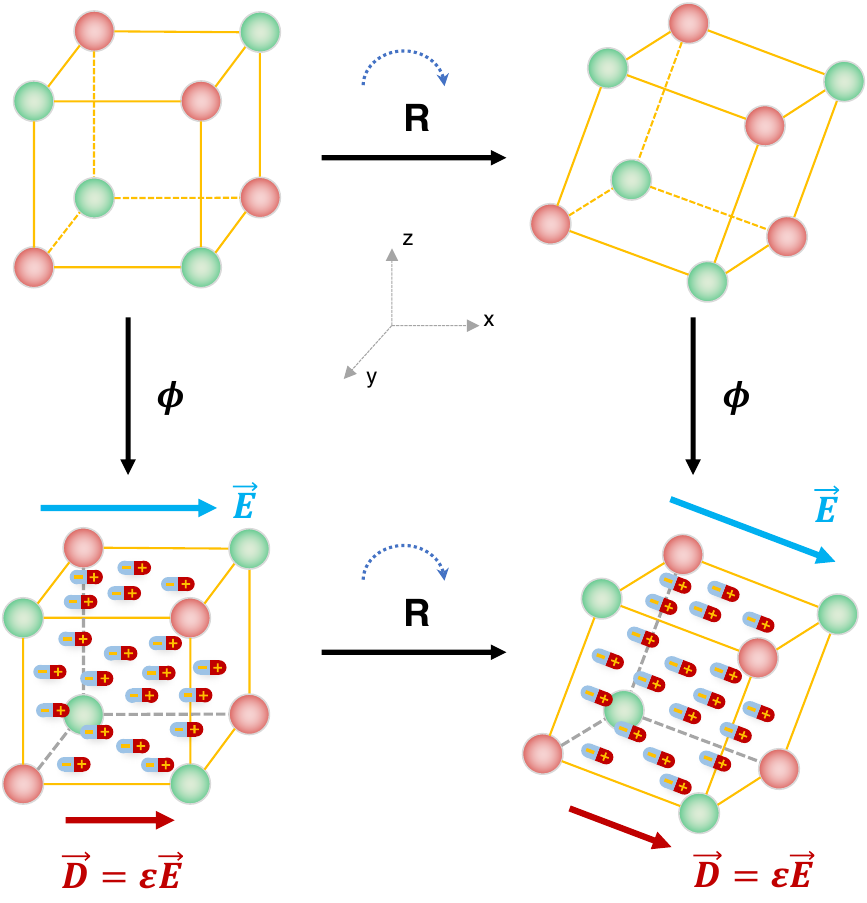}
	\caption{Rotational equivariance on dielectric constants of inorganic materials. $\protect\overrightarrow{E}$ is the external electric field, $\protect\overrightarrow{D}$ is the corresponding electric displacement, and $\boldsymbol{\varepsilon}$ is the $3\times3$ dielectric tensor.}
	\label{fig:equi}
\end{figure}

Atomic systems are usually represented with coordinates in 3D Euclidean space.
The same system could have transformations under the 3D symmetries: rotations, translations, and inversion.
Under orthogonal transformations in 3D space, certain properties of the system remain invariant, e.g., energy, 
meaning they do not change regardless of the rotation.
Equivariant properties, including forces, stresses, and dielectric tensors, 
are expected to transform accordingly with the system. In other words, as the system rotates or reflects, 
these properties undergo corresponding transformations to maintain their relationship with the system's new orientation.
Formally, we consider a function $\phi:X\rightarrow Y$ 
and a group transformation $G_T$ that acts on the vectors spaces $X$ and $Y$.
The function $\phi$ is equivariant to the transformation $G_T$ 
if the following condition holds for every element $x\in X$: $\phi(G_T(x))=G_T'(\phi(x))$, 
where $G_T(x)$ represents the group transformation applied to the input $x$,
and $G_T'(f(x))$ represents a related transformation applied to the output $\phi(x)$.
Here, we focus on the rotation/inversion equivariance of the second-rank dielectric tensor $\boldsymbol{\varepsilon}$ with respect to the input vector space $\mathcal{X}$ (Fig.\ref{fig:equi}),
whose equivariant condition is $R\Phi(\mathcal{X})R^T=\Phi(R\mathcal{X})$
where $R$ is the arbitrary orthogonal matrix and $\Phi$ is the neural networks to map $\mathcal{X}$ to $\boldsymbol{\varepsilon}$.

\subsection{Dielectric Tensor Prediction Covering 72 Elements}

\subsubsection{Data preparation}

\begin{figure}[tb]
	\centering
	\includegraphics[width=0.75\textwidth]{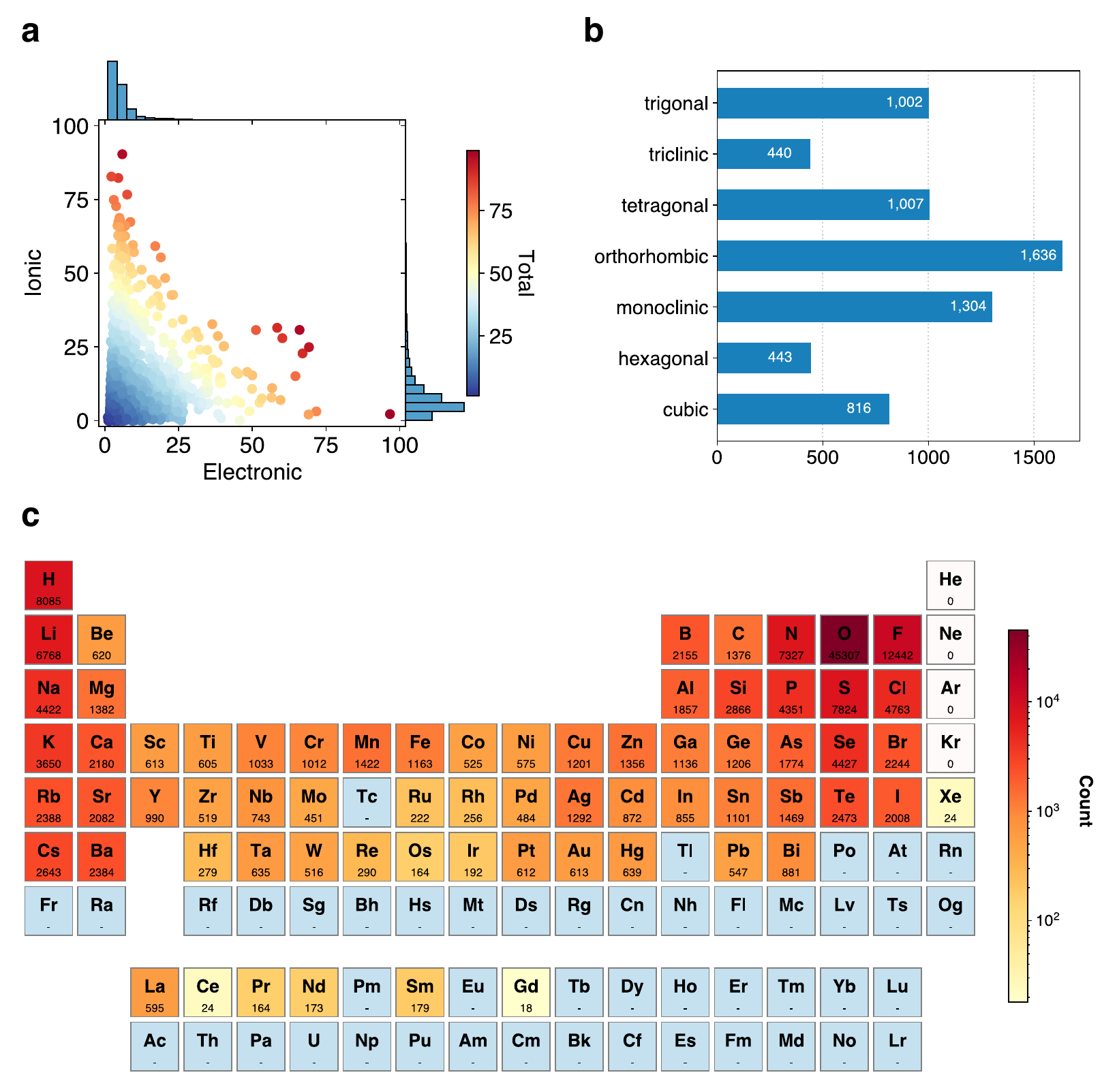}
	\caption{(a) $\varepsilon^\infty-\varepsilon^0$ map of structures in the dataset, colored by $\varepsilon$; (b)Counts of crystal system types in teh dataset; (c) Element counts for atoms in the structures with available dielectric properties on Materials Project. Elements under blue are not supported by \textit{PFP}. Elements under white are supported by \textit{PFP} but do not appear in the dataset.}
	\label{fig:data_dist}
\end{figure}

We leverage on one of the largest open databases of DFT-calculated crystal structure properties, Materials Project~\cite{jain2013commentary}, to obtain a wide range of training structures.
The dataset (MP v2023.11.1) contains 7277 relaxed materials along with calculated dielectric constants.
We clean up the dataset by restricting the element in total dielectric constants $\varepsilon_{ij} \in [-10, 100]$ for $\forall i,j$ 
and filtering out structures that contain elements not supported by \textit{PFP}.
Finally there are 6,648 structures retained for model training.
The data covers the electronic component $\varepsilon^\infty$, ionic component $\varepsilon^0$, and total component $\varepsilon$ with its scalar values in the range of $[1.0, 96.688]$, $[0.0, 90.338]$ and $[1.155, 98.889]$, respectively (Fig. \ref{fig:data_dist}(a)).
Fig. \ref{fig:data_dist}(b) exhibits that all seven types of crystal systems are included in our dateset.
Among them, the orthorhombic crystal system is the most prevalent, comprising 1,636 instances. 
In contrast, the triclinic and hexagonal systems are the least represented, with only 440 and 443 instances, respectively.
The version of \textit{PFP} architecture introduced in this work supports 72 elements in total, covering a wide range of elements across the periodic table. Excluding noble gases, He, Ne, Ar, and Kr, 68 elements are covered in the 6,648 available structures (Fig. \ref{fig:data_dist}(c)).

The above dataset was partitioned into training, validation, and test subsets in an 0.8:0.1:0.1 ratio for the experimental analysis. 
To mitigate the impact of random variation, each experiment was independently performed five times with distinct random data splits. 
The outcomes are presented as the mean and standard deviation of these five iterations.

\subsubsection{Performance of \textit{DTNet}}

\begin{table}[tb]
	\centering
	\caption{Number of independent components in the dielectric tensor for different crystal systems}
	\begin{tabular}{c c c c c c}
		\specialhline{2pt}
		Crystal & Number of independent elements & Dielectric tensor \\
		\hline
		\vspace{0.01cm}\\
		Cubic & 1 & $\begin{bmatrix}
			\varepsilon & 0 & 0\\
			0 & \varepsilon & 0\\
			0 & 0 & \varepsilon
		\end{bmatrix}$  \\
		\vspace{0.01cm}\\
		
		\hline
		\vspace{0.01cm}\\
		Tetragonal, Triagonal, Hexagonal & 2 & $\begin{bmatrix}
			\varepsilon_1 & 0 & 0\\
			0 & \varepsilon_1 & 0\\
			0 & 0 & \varepsilon_3
		\end{bmatrix}$  \\
		\vspace{0.01cm}\\
		
		\hline
		\vspace{0.01cm}\\
		Orthorhombic & 3 & $\begin{bmatrix}
			\varepsilon_1 & 0 & 0\\
			0 & \varepsilon_2 & 0\\
			0 & 0 & \varepsilon_3
		\end{bmatrix}$  \\
		\vspace{0.01cm}\\
		
		\hline
		\vspace{0.01cm}\\
		Monoclinic & 4 & $\begin{bmatrix}
			\varepsilon_{11} & 0 & \varepsilon_{13}\\
			0 & \varepsilon_2 & 0\\
			\varepsilon_{13} & 0 & \varepsilon_{33}
		\end{bmatrix}$  \\
		\vspace{0.01cm}\\
		
		\hline
		\vspace{0.01cm}\\
		Triclinic & 6 & $\begin{bmatrix}
			\varepsilon_{11} & \varepsilon_{12} & \varepsilon_{13}\\
			\varepsilon_{12} & \varepsilon_{22} & \varepsilon_{23}\\
			\varepsilon_{13} & \varepsilon_{23} & \varepsilon_{33}
		\end{bmatrix}$  \\
		\vspace{0.01cm}\\
		\specialhline{2pt}
	\end{tabular}
	\label{tab:independent}
\end{table}

\begin{figure}[htp]
	\centering
	\includegraphics[width=130mm]{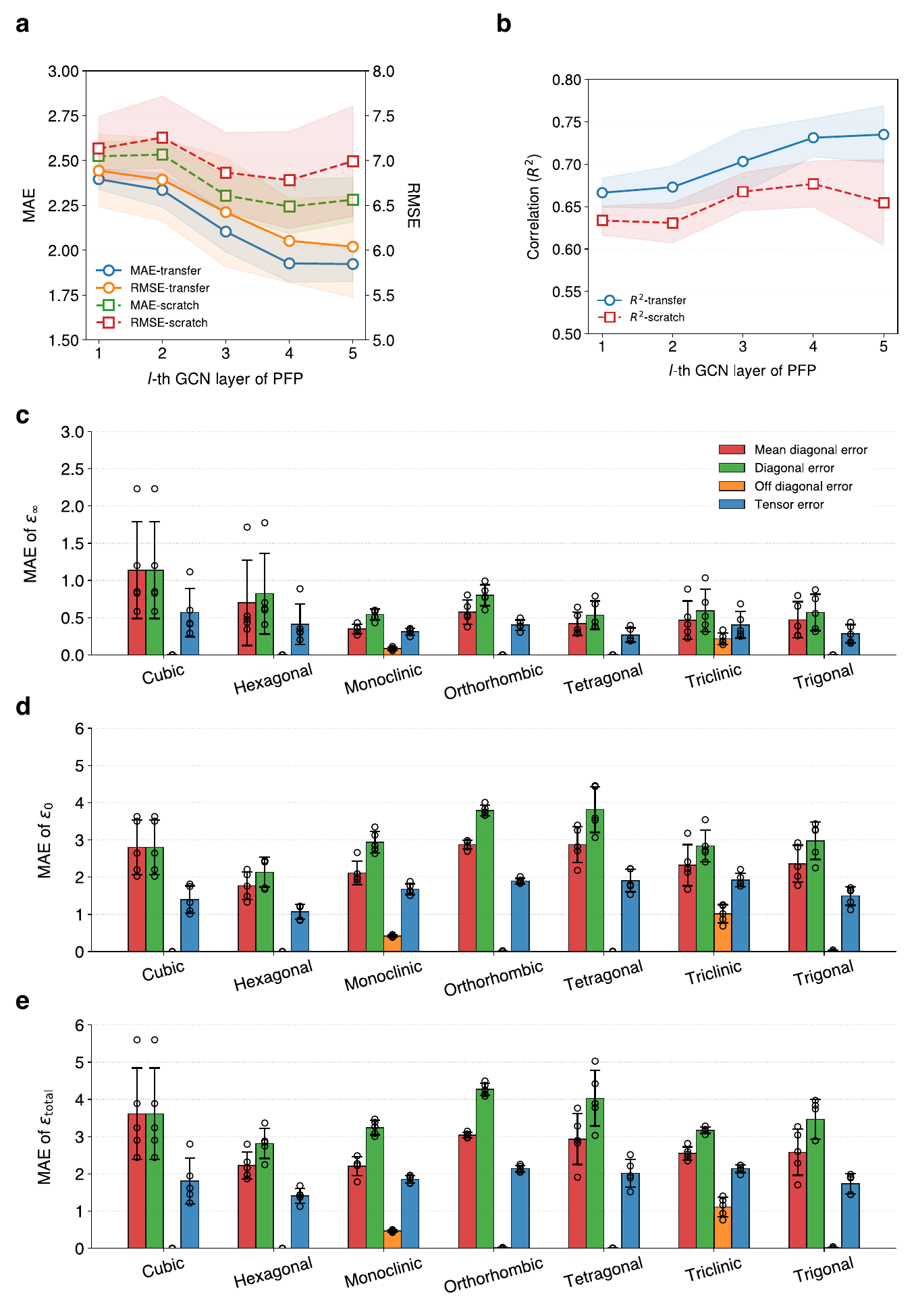}
	\caption{Performance of \textit{DTNet} models on dielectric data from Materials Project. (a-b) Plot of the mean error/correlation (points) with standard deviation (shadow) for 5 runs against the intermediate embedding generated from the \textit{PFP}-L\textsubscript{n} as \textit{DTNet} inputs, showing both transfer learning results and training from scratch results for comparison. (c-e) Multiple MAE metrics in the dielectric tensor for different crystal systems on the prediction of the electronic, ionic and total task.}
	\label{fig:pred}
\end{figure}

We start by investigating the appropriate intermediate GCN layer of \textit{PFP} as the input to \textit{DTNet}. 
Note that neuron parameters related to $\boldsymbol{a}_v^5$, $\boldsymbol{a}_t^5$ and $\boldsymbol{b}_v^5$ in the 5-th GCN layer are not trained in \textit{PFP} 
because only $\boldsymbol{a}_s^5$ and $\boldsymbol{b}_s^5$ are used as readout features. 
Hence, for the multi-order features generated from the \textit{PFP}-L\textsubscript{$n$}, we take $\boldsymbol{a}_s^n$, $\boldsymbol{a}_v^n$, $\boldsymbol{a}_t^n$, $\boldsymbol{b}_s^n$, and $\boldsymbol{b}_v^n$ for $1\leq n\leq 4$, 
and we take $\boldsymbol{a}_s^5$, $\boldsymbol{a}_v^4$, $\boldsymbol{a}_t^4$, $\boldsymbol{b}_s^5$, and $\boldsymbol{b}_s^4$ for the \textit{PFP}-L\textsubscript{5}.
We have performed hyperparameter optimization for the \textit{DTNet} architecture via the optuna framework~\cite{akiba2019optuna}.
The hyperparameter configurations are detailed in Method Section.
Finally, 2 equivariant blocks (see Fig. \ref{fig:model}(b)) are stacked and combined with the \textit{PFP}-L\textsubscript{$n$}.
For a batch of $B$ test structures, four mean absolute error (MAE) metrics for the symmetric dielectric tensor, 
the tensor $\mathrm{MAE_{ten}}$, mean diagonal $\mathrm{MAE_{mean-diag}}$, diagonal $\mathrm{MAE_{diag}}$, and off-diagonal $\mathrm{MAE_{off-diag}}$, are defined respectively as:
\begin{equation}
	\mathrm{MAE_{ten}} = \frac{1}{6B}\sum^B\sum_{1\leq i\leq 3, i\leq j \leq 3}\abs{\varepsilon_{ij}^{\mathrm{dft}}-\varepsilon_{ij}^{\mathrm{pred}}}
\end{equation}
\begin{equation}
	\mathrm{MAE_{mean-diag}} = \frac{1}{B}\sum^B\abs{\frac{1}{3}\sum_{1\leq i=j \leq 3}\varepsilon_{ij}^{\mathrm{dft}}-\frac{1}{3}\sum_{1\leq i=j \leq 3}\varepsilon_{ij}^{\mathrm{pred}}}
\end{equation}
\begin{equation}
	\mathrm{MAE_{diag}} = \frac{1}{3B}\sum^B\sum_{1\leq i=j \leq 3}\abs{\varepsilon_{ij}^{\mathrm{dft}}-\varepsilon_{ij}^{\mathrm{pred}}}
\end{equation}
\begin{equation}
	\mathrm{MAE_{off-diag}} = \frac{1}{3B}\sum^B\sum_{1\leq i \leq 3, i < j\leq 3}\abs{\varepsilon_{ij}^{\mathrm{dft}}-\varepsilon_{ij}^{\mathrm{pred}}}
\end{equation}
These MAE metrics help assess the accuracy of the model predictions for the dielectric tensor properties.
In this work, all models are trained using $\mathrm{MAE_{ten}}$ as the loss function.
Due to structural symmetry, the number of independent elements in the dielectric tensor varies according to system types (see Table\ref{tab:independent}).
Fig. \ref{fig:pred}(a-b) shows that generally the intermediate knowledge from the deeper GCN layer contributes more to the dielectric prediction.
When \textit{PFP}-L\textsubscript{4} and \textit{PFP}-L\textsubscript{5} are taken as the feature generator for the child model, lower $\mathrm{MAE_{ten}}$ prediction errors are observed on the test structures.
This could be explained by that the shallower layers only capture shorter-range local interactions.
Meanwhile, quite close accuracy between $n=4$ and $n=5$ is because only their scalar representations are distinct.
Considering the better computational efficiency, \textit{PFP}-L\textsubscript{4} is taken as the feature generator and is utilized for the subsequent study.
The comparison of error metrics between models utilizing transfer learning and those trained from scratch highlights the significance of incorporating compositional and structural encodings from \textit{PFP}.
Specifically, the use of pretrained \textit{PFP}-L\textsubscript{4} resulted in a reduction of 10.0\% for RMSE and 14.1\% in MAE compared to a \textit{PFP}-L\textsubscript{4} model with the same architecture but trained with all parameters initialized from scratch.
Fig. \ref{fig:pred}(c-e) showcases the performance of \textit{DTNet} on each crystal system for the prediction task of the electronic, ionic and total dielectric constants.
Three models \textit{DTNet}-$\varepsilon^\infty$, \textit{DTNet}-$\varepsilon^0$ and \textit{DTNet}-$\varepsilon$ are separately trained
with their $\mathrm{MAE_{ten}}, \mathrm{MAE_{mean-diag}}, \mathrm{MAE_{diag}}, \mathrm{MAE_{off-diag}}$ evaluated.
Dielectric tensors of cubic, hexagonal, orthorhombic, tetragonal and trigonal systems hold 0 on their off-diagonal elements due to geometric symmetry.
The zero off-diagonal errors for them
demonstrates the our model keeps equivariance for correct prediction of 0 on all their off-diagonal elements.
Additionally, the same $\mathrm{MAE_{ten}}$ and $\mathrm{MAE_{mean-diag}}$ of the cubic system further confirm
the output equivariance is guaranteed due to its symmetric structure along all three directions in the Euclidean space.
The model performs uniformly for different crystal systems according to all $\mathrm{MAE_{ten}}$ in the range of $[0.27, 0.57]$ for $\boldsymbol{\varepsilon^\infty}$, $[1.07, 1.93]$ for $\boldsymbol{\varepsilon^0}$ and $[1.41, 2.14]$ for $\boldsymbol{\varepsilon}$.

\begin{figure}[tb]
	\centering
	\includegraphics[width=0.8\textwidth]{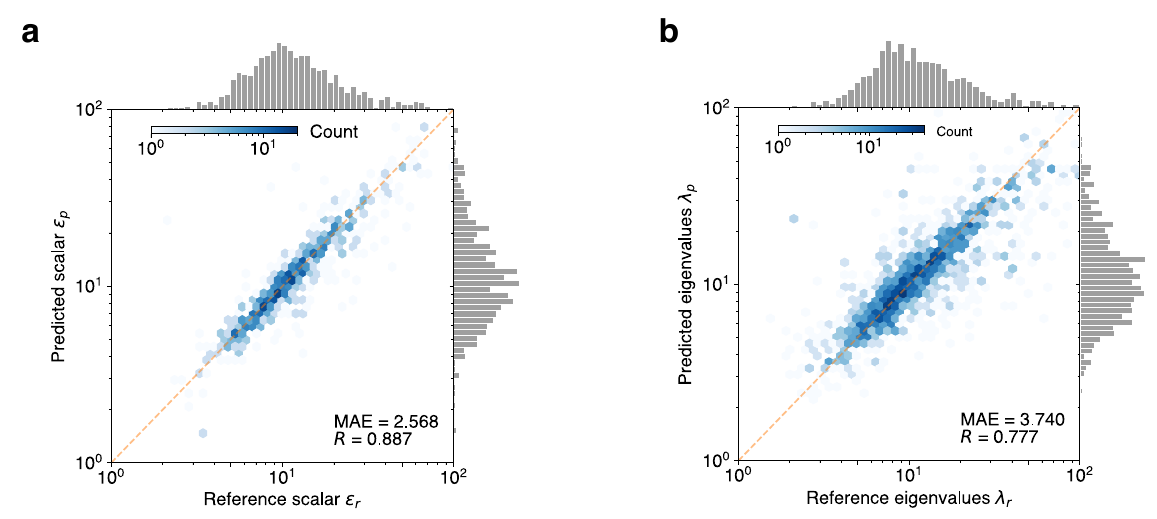}
	\caption{
            Performance of \textit{DTNet} on the test set of our refined MP dataset for two distinct tasks: (a) polycrystalline dielectric constants and (b) tensor eigenvalues.
            }
	\label{fig:tensor_pred}
\end{figure}

Fig.~\ref{fig:tensor_pred}(a-b) illustrates the performance of \textit{DTNet} by calculating the polycrystalline dielectric constants $\varepsilon$ and tensor eigenvalues $\lambda_i$ from the predicted dielectric tensors, with the eigenvalues being crucial for evaluating anisotropy.
The ground trues of $\varepsilon$ and $\lambda_i$ in the test set fall within the range of [2.129, 98.889] and [1.460, 98.889], respectively.
\textit{DTNet} provides predictions that closely match DFT reference values, with MAE of 2.568 for $\varepsilon$ and 3.740 for $\lambda_i$, within a prediction range of 96.760 and 97.429, respectively.
To further validate the accuracy of our method, we evaluated its performance on the dielectric task using the Matbench dataset~\cite{dunn2020benchmarking}, which is a subset of structures from the Materials Project.
In Matbench, we assessed our method by calculating refractive index $\eta$ from the dielectric tensors predicted by \textit{DTNet}.

\begin{figure}[tb]
	\centering
	\includegraphics[width=0.8\textwidth]{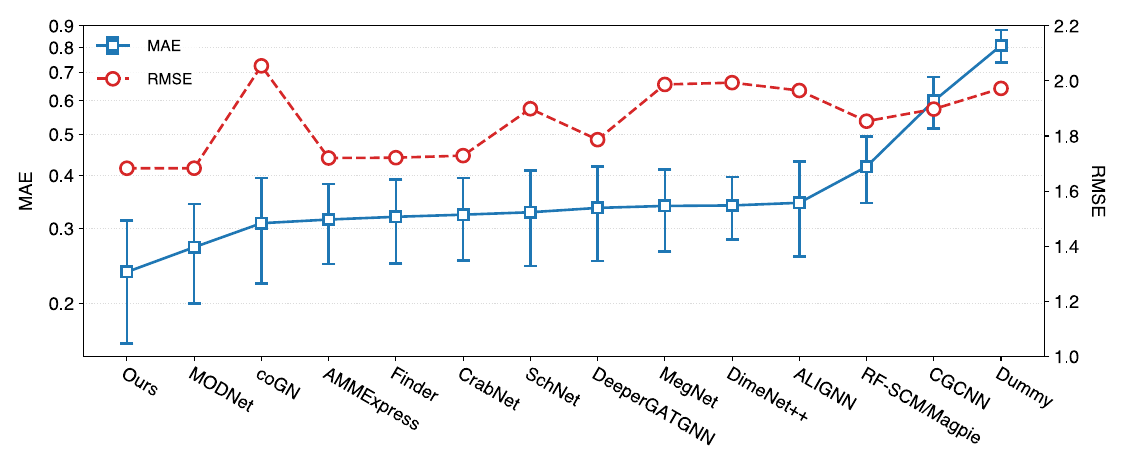}
	\caption{
        Benchmarking our approach against other algorithms on the Matbench dataset. 
		All results are sourced from the official Matbench leaderboards. 
		For the MODNet, coGN, and Finder models, we present only their best reported results.}
	\label{fig:matbench}
\end{figure}

Fig.~\ref{fig:matbench} compares the mean MAEs with error bars representing standard deviations and the mean RMSEs across 5 runs, as defined by the Matbench pipeline.
\textit{DTNet} surpasses all other methods published on the official Matbench leaderboards for the dielectric task, achieving an average MAE of 0.2371 with a standard deviation of 0.0765, and an RMSE of 1.6830 over the 5-fold test data.
In comparison, the top-performing model MODNet~\cite{de2021materials} reported on Matbanch has an MAE of 0.2711 with a standard deviation of 0.0714 and an RMSE of 1.6832.

\subsubsection{Model comparison for dielectric tensor prediction}

At the time of conducting this work, there were limited open-source works that discussed the prediction of tensorial properties of materials.
The models of \textit{PaiNN}~\cite{schutt2021equivariant}, \textit{M3GNet}~\cite{chen2022universal} and \textit{MatTen}~\cite{wen2023universal}
are chosen for comparison in this work.
They use different approaches to produce rank-2 equivariant tensors by incorporation of equivariant atomwise representations,
partial derivative with respect to the Cartesian tensor,
and spherical harmonics expansion, respectively.
\textit{PaiNN} encodes each atom $i$ with an invariant representation $\boldsymbol{s}_i$
and an equivariant vector representation $\boldsymbol{v}_i$.
Incorporating $\boldsymbol{v}_i$ has been demonstrated to be effective for structure recognition through ablation studies.
The dielectric tensor in \textit{PaiNN} is constructed using a rank-1 tensor decomposition, 
and the atom positions $\boldsymbol{r}_i$ are incorporated to hold the global structure information:
\begin{equation}
	\boldsymbol{\varepsilon} = \sum_{i=1}^{N^a}s(\boldsymbol{s}_i)\boldsymbol{I}_3+\nu(\boldsymbol{v}_i)\otimes\boldsymbol{r}_i+\boldsymbol{r}_i\otimes\nu(\boldsymbol{v}_i)
\end{equation}
where $\boldsymbol{I}_3$ is an identity matrix with size of three, $s$ and $\nu$ are invariant and equivariant nonlinearities, respectively.
The invariant \textit{M3GNet} model is able to generate equivariant $3\times 3$ tensor from 
the partial derivatives of its scalar output using the lattice matrix information.
This scalar output is not directly trained as a target variable, 
and instead we train the model based on the loss computed from the derivative values of this output. 
Specifically, we calculate the loss of $\mathrm{MAE}_{ten}$ between these derivative results and labeled targets.
After training, these derivative values are utilized for dielectric prediction.
In MatTen,
the atom feature contains a set of high-rank tensors produced by spherical harmonics,
namely, a geometric object consisting of a direct sum of irreducible representations of the SO(3) rotation group.
Its interaction blocks follow the design of Tensor Field Network~\cite{thomas2018tensor} and NequIP~\cite{batzner20223}.
The second-order features of all atoms are ultimately aggregated as the prediction of dielectric tensor.
To investigate the necessity of including edge-related representations generated by \textit{PFP}-L\textsubscript{4}, we trained two models, \textit{DTNet} and \textit{DTNet}-simple.
\textit{DTNet} leverages all latent embeddings $(\boldsymbol{a}_s^n, \boldsymbol{a}_v^n, \boldsymbol{a}_t^n, \boldsymbol{b}_s^n, \boldsymbol{b}_v^n)$, while \textit{DTNet}-simple only utilizes $(\boldsymbol{a}_s^n, \boldsymbol{a}_v^n, \boldsymbol{a}_t^n)$ and the layers related to Eq.\ref{eq: edge scalar} and Eq.\ref{eq: edge vector} are removed.

Table \ref{tab:pred_system} provides a comparison of the three aforementioned models with \textit{DTNet} and \textit{DTNet}-simple.
In the prediction task across different systems, the models are trained on the entire training data, while the test structures are categorized according to crystal systems.
It is shown that \textit{DTNet} performs best in 22 among all 24 tasks in comparison with all models.
The aggregation of potential edge information showcases significant improvements in accuracy,
as evidenced by \textit{DTNet} beating \textit{DTNet}-simple in 18 out of the 24 tasks.
It should be noted that \textit{DTNet}-simple still gives much better accuracy than the other 3 models,
outperforming 22 tasks. 
Only two tasks lost to \textit{M3GNet} when making ionic and total dielectric tensor predictions on triclinic systems.
Hence, the latent edge scalar and vector features provide additional useful geometric information which benefits the dielectric prediction.

\begin{table}[tb]
	\centering
	\caption{Performance comparison of models across all crystal systems.
		Models are trained separately for 3 tasks: the electronic component, ionic component and total dielectric constants.
		The results are shown by the $\mathrm{MAE_{ten}}$ and standard deviation of 5 runs using different random split.
		The 5 trained \textit{DTNet} models will be employed as ensemble models for virtual screening in the next section.  
	}
	\begin{tabular}{c c c c c c}
		\specialhline{2pt}
		Models & \textit{DTNet} & \textit{DTNet}-simple & \textit{PaiNN} & \textit{M3GNet} & \textit{MatTen}\\
		\hline
		Task & \multicolumn{5}{c}{Electronic ($\varepsilon_{\infty}$)}\\
		\hline
		Cubic          & $\mathbf{0.570\pm0.325}$  & $0.714\pm0.449$           & $1.811\pm0.900$ & $0.971\pm0.498$         & $1.012\pm0.340$ \\
		Hexagonal      & $\mathbf{0.412\pm0.272}$  & $0.430\pm0.260$           & $0.893\pm0.537$ & $0.469\pm0.310$         & $0.548\pm0.274$ \\
		Monoclinic     & $\mathbf{0.314\pm0.045}$  & $0.319\pm0.047$           & $0.515\pm0.115$ & $0.382\pm0.070$         & $0.447\pm0.025$ \\
		Orthorhombic   & $0.401\pm0.071$           & $\mathbf{0.391\pm0.088}$  & $0.749\pm0.200$ & $0.420\pm0.059$         & $0.523\pm0.086$ \\
		Tetragonal     & $\mathbf{0.268\pm0.095}$  & $0.268\pm0.095$           & $0.665\pm0.224$ & $0.456\pm0.191$         & $0.415\pm0.092$ \\
		Triclinic      & $0.405\pm0.179$           & $\mathbf{0.404\pm0.170}$  & $0.722\pm0.282$ & $0.431\pm0.079$         & $0.564\pm0.214$ \\
		Trigonal       & $\mathbf{0.285\pm0.122}$  & $0.311\pm0.133$           & $0.737\pm0.327$ & $0.371\pm0.103$         & $0.494\pm0.191$ \\
		All            & $\mathbf{0.368\pm0.063}$  & $0.391\pm0.063$           & $0.836\pm0.125$ & $0.489\pm0.085$         & $0.554\pm0.049$ \\
		\hline
		Task & \multicolumn{5}{c}{Ionic ($\varepsilon_0$)}\\
		\hline
		Cubic          & $1.400\pm0.368$           & $\mathbf{1.383\pm0.329}$  & $3.217\pm1.025$ & $1.901\pm0.321$         & $1.819\pm0.621$ \\
		Hexagonal      & $1.068\pm0.201$           & $\mathbf{1.028\pm0.121}$  & $2.168\pm0.479$ & $1.359\pm0.206$         & $1.572\pm0.308$ \\
		Monoclinic     & $\mathbf{1.679\pm0.147}$  & $1.711\pm0.130$           & $2.351\pm0.272$ & $1.827\pm0.175$         & $1.982\pm0.178$ \\
		Orthorhombic   & $\mathbf{1.898\pm0.074}$  & $1.915\pm0.114$           & $3.120\pm0.171$ & $1.957\pm0.150$         & $2.156\pm0.173$ \\
		Tetragonal     & $1.910\pm0.305$           & $\mathbf{1.867\pm0.348}$  & $3.14\pm0.398$  & $2.058\pm0.192$         & $2.092\pm0.265$ \\
		Triclinic      & $1.926\pm0.179$           & $1.901\pm0.144$           & $2.523\pm0.229$ & $\mathbf{1.783\pm0.267}$& $2.172\pm0.201$ \\
		Trigonal       & $\mathbf{1.494\pm0.253}$  & $1.571\pm0.315$           & $2.745\pm0.600$ & $1.742\pm0.230$         & $1.769\pm0.333$ \\
		All            & $\mathbf{1.677\pm0.050}$  & $1.685\pm0.049$           & $2.823\pm0.180$ & $1.862\pm0.131$         & $1.969\pm0.052$ \\
		\hline
		Task & \multicolumn{5}{c}{Total ($\varepsilon$)}\\
		\hline
		Cubic          & $\mathbf{1.808\pm0.615}$  & $1.898\pm0.621$           & $4.834\pm1.443$ & $2.627\pm0.702$         & $2.548\pm0.845$ \\
		Hexagonal      & $\mathbf{1.409\pm0.199}$  & $1.419\pm0.271$           & $3.151\pm0.498$ & $1.671\pm0.557$         & $1.954\pm0.266$ \\
		Monoclinic     & $\mathbf{1.854\pm0.098}$  & $1.890\pm0.148$           & $2.784\pm0.344$ & $2.047\pm0.160$         & $2.267\pm0.184$ \\
		Orthorhombic   & $\mathbf{2.141\pm0.089}$  & $2.178\pm0.068$           & $3.535\pm0.152$ & $2.228\pm0.153$         & $2.461\pm0.180$ \\
		Tetragonal     & $\mathbf{2.018\pm0.373}$  & $2.036\pm0.284$           & $3.325\pm0.544$ & $2.403\pm0.345$         & $2.351\pm0.434$ \\
		Triclinic      & $2.142\pm0.105$           & $2.218\pm0.181$           & $3.103\pm0.624$ & $\mathbf{1.906\pm0.284}$& $2.488\pm0.136$ \\
		Trigonal       & $1.739\pm0.269$           & $\mathbf{1.624\pm0.280}$  & $3.265\pm0.709$ & $1.984\pm0.259$         & $2.040\pm0.371$ \\
		All            & $\mathbf{1.911\pm0.094}$  & $1.927\pm0.104$           & $3.427\pm0.265$ & $2.185\pm0.114$         & $2.320\pm0.153$ \\
		\specialhline{2pt}
	\end{tabular}
	\label{tab:pred_system}
\end{table}

\subsection{Discovery of Dielectric Materials}
\subsubsection{Preparation of candidate structures}
The \textit{DTNet} model has been proven effective predicting dielectric tensors which account for the direction and magnitude of the electric field.
To prepare the candidate set for screening, we downloaded 14,375 non-metal materials from the MP database.
These materials are specifically selected based on the energy above convex hull $E_{\mathrm{hull}}=0$ to estimate their thermodynamic stability,
so that only stable materials are included in the candidate set.
The candidates have unknown dielectric constant properties, and there is no overlap between this candidate set and our training dataset.
We aim to leverage our models for virtual screening on all these candidates to identify potential materials for high-dielectric materials and highly anisotropic materials, incorporating both electronic and ionic contributions.
To improve accuracy and assess model uncertainty, we use ensemble results from the five \textit{DTNet} models trained in the previous section. 
The expected property $\mu(\chi)$ is estimated by averaging the predictions, while the uncertainty $\sigma(\chi)$ is quantified by the standard deviations.
We conducted iterative active learning to identify top candidates through virtual screening.
See details in the Method section.

\subsubsection{High-dielectric materials}
Dielectrics are crucial functional materials for microelectronic device manufacturing.
Materials with high dielectric constants have the potential to enable higher energy density storage, 
leading to performance enhancement and device miniaturization~\cite{qu2020high}.
Meanwhile, large band gaps ($E_g$) are also desirable as they prevent leakage currents when materials are exposed to large electric fields, 
especially at nanoscale thicknesses.
Hence, novel dielectric materials with both high dielectric constants $\varepsilon$ and large band gaps $E_g$ are highly preferred for such applications.
To represent the capacity of electric energy storage, dielectric tensors $\boldsymbol{\varepsilon}$ can be simplified to scalar quantities by Eq.\ref{eq:scalar dielectric}.

\begin{figure}[tb]
	\centering
	\includegraphics[width=0.9\textwidth]{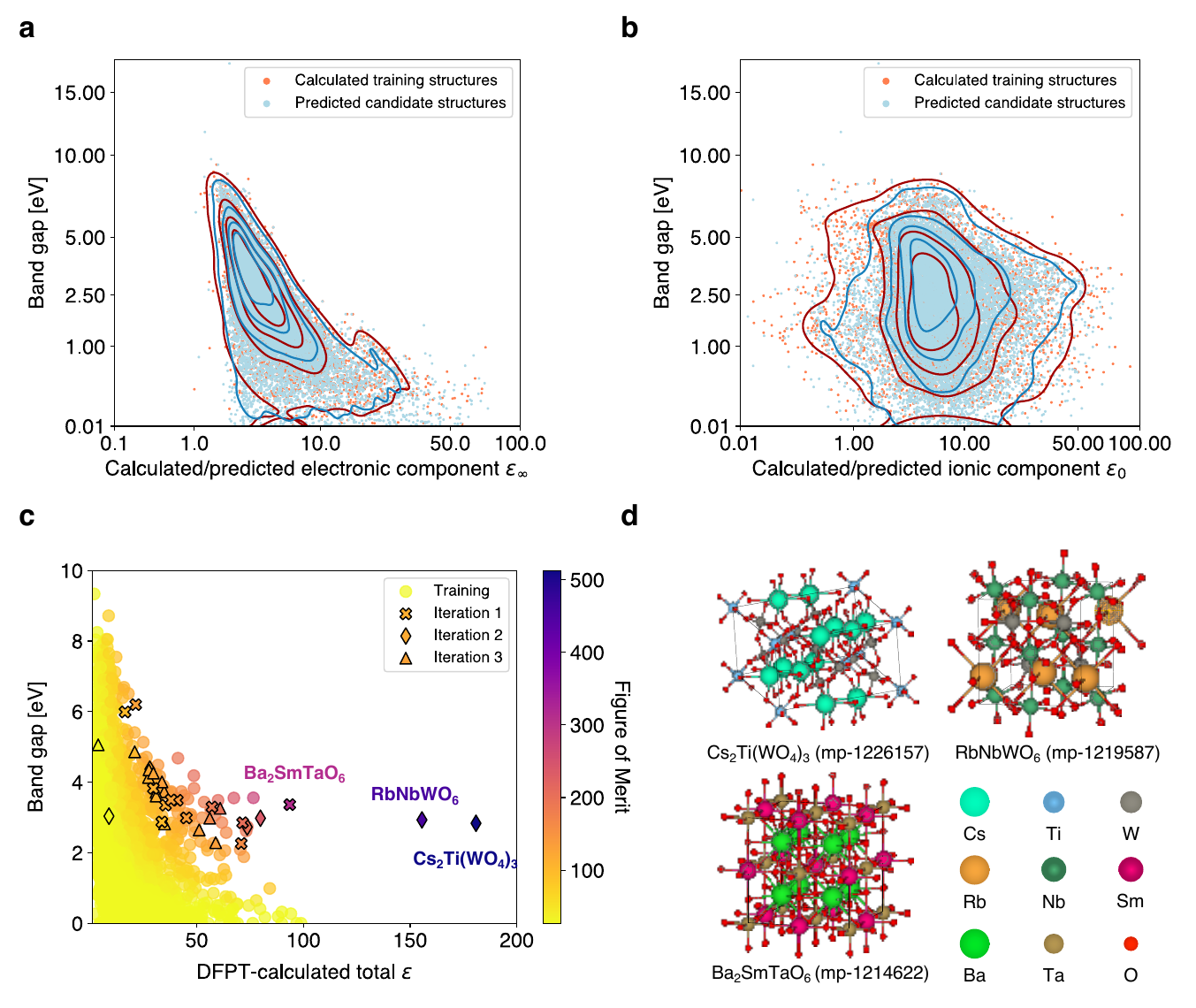}
	\caption{(a-b) The joint distribution of band gaps $E_g$ and electronic dielectric constants $\varepsilon^\infty$ (a) 
		or ionic dielectric constants $\varepsilon^0$ (b) for 6,648 training structures and 14,375 stable candidate structures.
		$\varepsilon^\infty$ and $\varepsilon^0$ of training structures and $E_g$ of all structures are obtained from 
		computational results in MP, while $\varepsilon^\infty$ and $\varepsilon^0$ of candidate structures are predicted by \textit{DTNet-$\varepsilon^\infty/\varepsilon^0$}.
		The 5-level densities of data points are drawn in contours using the kernel density estimation for structures both in the candidate set and original set.
		(c) DFT-calculated $E_g-\varepsilon$ data points colored by their figure of merit defined by $E_g\cdot\varepsilon$
		for 4,146 stable structures in training data and 35 new stable structures screened by \textit{DTNet} across 3 iterations.
		(d) Structure visualization of the top 3 materials by FOM as validated by DFPT calculations.}
	\label{fig:results_bandgap}
\end{figure}

It is known that there exists inverse relationship between the electronic part of dielectric constants $\varepsilon^{\infty}$ and band gaps $E_g$.
On the other hand, the ionic contribution $\varepsilon^0$ does not show a clear correlation with $E_g$.
The theoretical background of this phenomenon is explained in reference~\cite{takahashi2020machine}.
This trend is also observed in the materials present in the training set as shown in Fig. \ref{fig:results_bandgap}(a) and Fig. \ref{fig:results_bandgap}(b).
To showcase the capability of our model in capturing such latent quantum mechanism knowledge underlying chemical structures,
we employ the \textit{DTNet-$\varepsilon^\infty$} and \textit{DTNet-$\varepsilon^0$} models to predict the two dielectric constant components
for the 14,375 candidates, respectively.
The density of data points are drawn in contours for structures in the training set (red) and candidate set (blue) using the kernel density estimation.
Despite the fact that the model does not learn explicitly from band gap data, 
the distribution of model-predicted electronic dielectric constants of the candidates and their DFPT-computed band gaps from MP indicates the model's capacity to comprehend this negative relationship between the two properties through materials' graph representations.
This phenomenon can be explained by the interplay between electronic polarization and electronic excitations. 
Materials with smaller band gaps allow for easier excitation of electrons to higher energy levels, leading to higher electronic polarizability and larger electronic dielectric constants. 
Conversely, materials with larger band gaps hinder electron excitation, resulting in lower electronic dielectric constants ~\cite{pilania2013accelerating}.
In contrast to the case of $\varepsilon^{\infty}$, 
no discernible correlation is observed between \textit{DTNet-$\varepsilon^0$}-predicted ionic part and band gaps.
From a microscopic perspective, the ionic dielectric constant is influenced by the displacement of cations and anions from their equilibrium positions under the influence of electric fields. 
The strength of bonds or phonon frequencies is highly sensitive to variations in bond lengths. 
Consequently, materials with similar compositions can exhibit a wide range of $\varepsilon^0$ due to differences in bond lengths ~\cite{lee2018high}.
The weak constraint of $E_g$ to $\varepsilon^0$ along with the broad value spectrum of $\varepsilon^0$ present 
the opportunity for the exploration and discovery of novel dielectric materials 
characterized by both high $\varepsilon$ and $E_g$.

We conducted the virtual screening by employing \textit{DTNet}-$\varepsilon$ to predict total dielectric constant scalar $\varepsilon$ for all structures in the candidate set.
Following the previous work~\cite{lee2018high}, $E_g\cdot\varepsilon$ is assigned as the figure of merit (FOM) to quantify the performance of dielectrics,
since $E_g$ and $\varepsilon$ are approximately proportional to the logarithm of the leakage current density~\cite{yim2015novel}.
To validate the promising materials identified by \textit{DTNet}, 
we conducted a three-round active exploration, with each round proposing 20 top candidates for validation.
The details of virtual screening and DFPT configurations are available in the Method section.
Finally, totally 35 out of the 60 structures across 3 iterations are successfully converged through first-principle calculations.

Fig.~\ref{fig:results_bandgap}(c) illustrates the distribution of the figure of merit (FOM) for stable materials in both the training set and screened candidate set.
The thermaldynamically stable material with the highest FOM in the training dataset is Sn\textsubscript{2}ClF\textsubscript{3} (mp-29252, $E_g=3.56 \mathrm{eV}, \varepsilon^\infty=3.81, \varepsilon^0=72.77, \varepsilon=76.55$).
We successfully identified 3 new materials,
Cs\textsubscript{2}Ti(WO\textsubscript{4})\textsubscript{3} (mp-1226157, $E_g=2.83 \mathrm{eV}, \varepsilon^\infty=5.34, \varepsilon^0=175.55, \varepsilon=180.89$),
RbNbWO\textsubscript{6} (mp-1219587, $E_g=2.93 \mathrm{eV}, \varepsilon^\infty=5.27, \varepsilon^0=150.30, \varepsilon=155.57$) and
Ba\textsubscript{2}SmTaO\textsubscript{6} (mp-1214622, $E_g=3.36 \mathrm{eV}, \varepsilon^\infty=4.85, \varepsilon^0=88.96, \varepsilon=93.81$),
with structures visualized in Fig. \ref{fig:results_bandgap}(d).
Their FOM scores are higher than any calculated stable structures in our training database.

\subsubsection{Highly anisotropic dielectrics}
Crystalline dielectrics can exhibit anisotropic dielectric properties \textit{w.r.t.} the direction of the applied electric field due to their cell structures and atomic arrangements.
Crystals with highly anisotropic dielectric properties can demonstrate unique characteristics, such as enhanced piezoelectricity, ferroelectricity~\cite{newnham2005properties}, and optical performance~\cite{tudi2022potential}.
Lou et al. explored the discovery of crystals with high anisotropy in high-frequency dielectric tensors, focusing solely on the electronic contribution~\cite{lou2024discovery}.
In this task, we tackle the more challenging task of discovering highly anisotropic static dielectric tensors, incorporating both electronic and ionic contributions, to achieve significantly greater anisotropy under the static external electric field.

\begin{figure}[tb]
	\centering
	\includegraphics[width=0.8\textwidth]{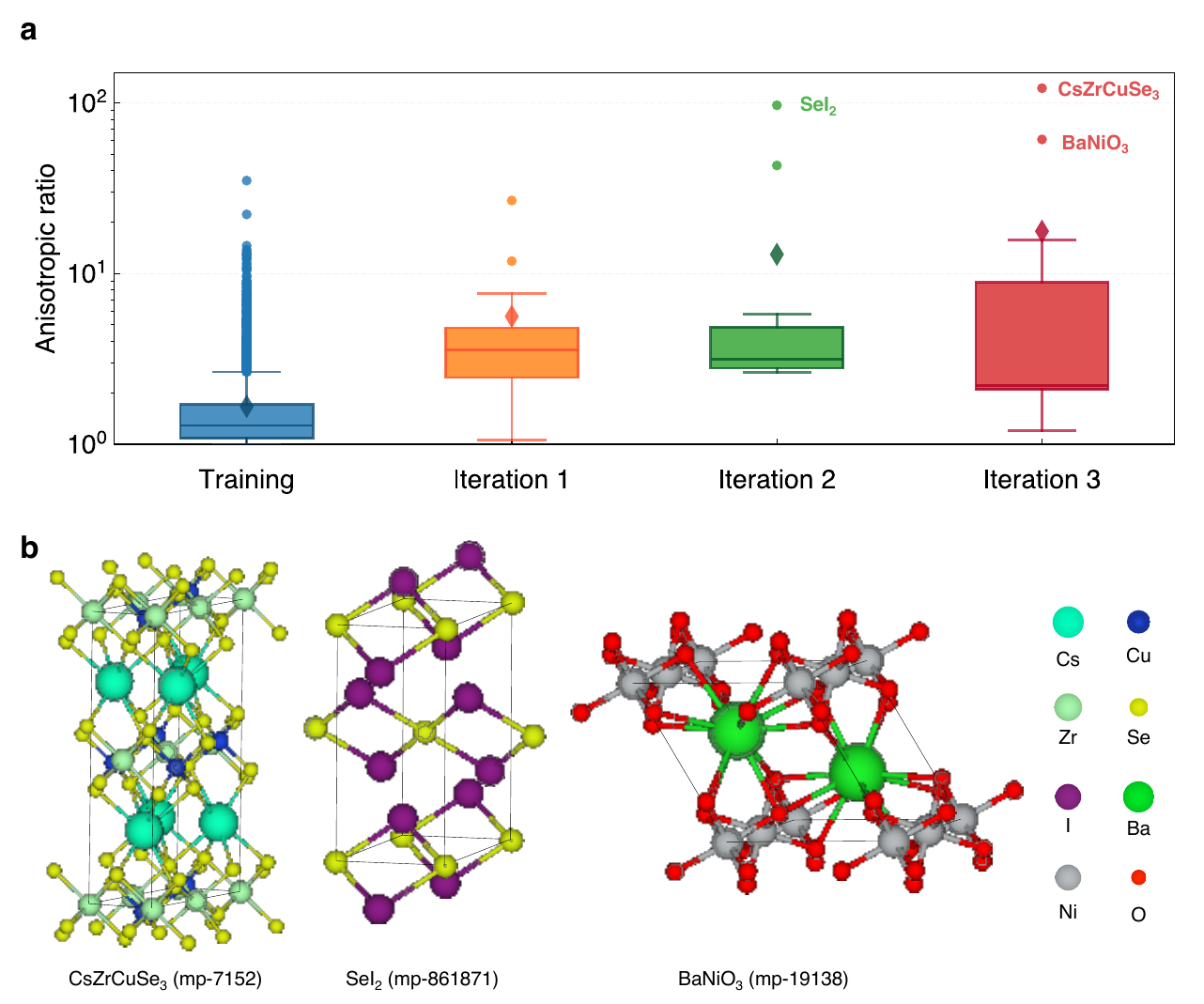}
	\caption{(a) Comparison of anisotropic distributions of 4,146 stable structures in training data and 41 new stable structures screened by \textit{DTNet} over 3 iterations.
		(d) Structure visualization of the top 3 materials by anisotropic ratios as validated by DFPT calculations.}
	\label{fig:results_aniso}
\end{figure}

Given the limited number of training samples with high anisotropy, we conducted a 3-round active learning for virtual screening using the same strategy to more effectively identify marginal candidates exhibiting high anisotropic ratio defined by $\alpha_r=\lambda_\textrm{max}/\lambda_\textrm{min}$~\cite{lou2024discovery}.
All molecular crystalline candidates were excluded prior to virtual screening due to their typically high anisotropy and potential instability under the conditions of interest.
For example, we evaluated the molecular crystal BrCl (mp-1066781). Despite exhibiting a high total anisotropy of $\alpha_r=26.774$, its melting point is as low as $-66^\circ C$, making it unstable under room temperature conditions.
Consequently, 41 out of the 60 screened structures were successfully validated via DFPT computations.
Fig.~\ref{fig:results_aniso}(a) presents the distribution of anisotropic ratios $\alpha_r$.
Notably, only 17 structures in the training dataset exhibit $\alpha_r$ exceeding 10, with the highest of 35.027.
In contrast to the average $\alpha_r$ of 1.665 observed in stable crystals within the dataset, the three iterative rounds produced significantly improved average $\alpha_r$ of 5.608, 12.946, and 17.687, respectively.
The top three DFPT-validated candidates were identified during the second and third iterations, with their structures visualized in Fig.~\ref{fig:results_aniso}(b).
Notably, CsZrCuSe\textsubscript{3} and BaNiO\textsubscript{3} are 3D materials, which may offer greater robustness and suitability for industrial applications compared to the 1D material SeI\textsubscript{2}.
The ionic components significantly contribute to the anisotropy in all three materials.
For CsZrCuSe\textsubscript{3}, the eigenvalue ranges for the electronic, ionic, and total contributions are [5.152, 12.851], [4.306, 1141.694], and [9.458, 1152.884], respectively, achieving electronic, ionic and total ratios of 2.494, 265.140 and 128.890, respectively.
The anisotropic ratios for SeI\textsubscript{2} are 2.745 (electronic), 316.090 (ionic), and 96.763 (total), and for BaNiO\textsubscript{3} they are 1.749 (electronic), 92.904 (ionic), and 61.026 (total).
Compared to the largest $\alpha_r=35.027$ in the training set (BrNO\textsubscript{3}, mp-29526), CsZrCuSe\textsubscript{3} is 93.863 larger and 2.680 times greater.
A recent research has also highlighted the excellent optoelectronic and thermoelectric properties of CsZrCuSe\textsubscript{3}~\cite{goumri2024unraveling}, underscoring its great potential for practical applications.

\section{Discussion}

Transfer learning is the most popular approach to tackle the data bottleneck which is common in materials science.
Previous studies have utilized transfer learning in GNN models, 
both for the same properties and across different properties, to demonstrate enhanced accuracy in materials science~\cite{jha2019enhancing, hutchinson2017overcoming, mao2023ai}.
For example, Chen et al. developed the AtomSets framework~\cite{chen2021atomsets}, 
which employs a straightforward model architecture
to process structure-embedded scalar information read from a pretrained MEGNet model~\cite{chen2019graph},
enabling high accuracy predictions with limited datasets.
However, the feasibility of transfer learning for properties across different tensorial orders has been less explored.
Our study indicates that more than scalar features, it is also possible to transfer
higher-rank equivariant compositional and structural information 
from a pretrained EGNN energy model to a different tensorial property prediction
with a small amount of available data, while maintaining the equivariance.
These equivariant representations may hold shared structural or even electronic-level information,
since the \textit{PFP} model aims to emulate iterative electronic relaxation.
This is evident through the accurate predictions our model achieved for both electronic and ionic dielectric constants, which are produced by different microscopic mechanisms.
Furthermore, the pretrained \textit{PFP} serves as an encoder that integrates electron density information,
assisting in comprehending the electronic-level relationship between the $E_g$ and $\varepsilon^\infty$.
Apart from dielectric constants, several works attempted to predict tensorial properties of materials
such as the polarizability~\cite{wilkins2019accurate}, quadrupole~\cite{thurlemann2022learning}, by using equivariant architectures.
The accuracy of such tasks could potentially be improved by fixing 
parameters in the shallower layers of an EGNN model 
pretrained on commonly computed properties like energy.

The \textit{PFP} model is pretrained on over 22 million structures covering the diverse chemical space.
It is efficient and expressive to take a a pretrained \textit{PFP} as the parent model
to generate locally-interacted features as the input to a much simpler child model.
In the current readout architecture, it consists of only 0.6 million trainable parameters
for the downstream tasks of dielectric constant prediction.
The model performing best on the validation accuracy appears at 
the average epoch of 263 out of 5 training runs for total dielectric constants.
By stopping the training after 200 epochs without improvement in validation accuracy,
the model training can be efficiently completed within approximately 2 hours 
using about 6GB memory on an Nvidia Tesla-V100 GPU.
We proved that such efficient training on encodings from the parent model can bring a promising property predictor even on a small data region as demonstrated in Table \ref{tab:pred_system}.
The incorporation of second-order equivariant features in our model may also contribute to high data efficiency as the symmetry of the dielectric tensors is automatically guaranteed due to the model architecture.
Based on only 6.6k sample structures with labeled tensor containing up to 6 independent elements from MP,
our model achieved a mean tensor prediction error of $0.368$ for $\boldsymbol{\varepsilon}^\infty$, $1.677$ for $\boldsymbol{\varepsilon}^0$ and $1.911$ for $\boldsymbol{\varepsilon}$.

Thanks to widely-supported elements in the pretrained \textit{PFP}, our model can also give relatively good prediction on rare earth elements, e.g., the Sm element in Ba\textsubscript{2}SmTaO\textsubscript{6}.
This compound emerged as one of the top high-dielectric candidates identified from a dataset of over 14,000 structures, based directly on its structural graph.
The rare earth elements have diverse industrial applications such as glass, lights, magnets, batteries, and catalytic converters and so on~\cite{balaram2019rare}, due to their unique physical and chemical properties.
Our model has the potential to make predictions about properties concerning rare elements, which typically suffer from limited data availability.

In this study, we demonstrate the applicability of our model to two design challenges, utilizing different metrics derived from the predicted dielectric tensors with high accuracy.
Accurate predictions of dielectric constants as tensors can contribute to understanding 
material behavior in the presence of electric fields and their interactions with other materials.
This knowledge can facilitate the development of new materials with tailored properties for specific applications,
such as energy storage, signal transmission, electronic component design, and cutting-edge fields such as dark matter detection~\cite{coskuner2021directional}.

\section{Conclusion}

An equivariant model is introduced to predict the second-order tensorial property, dielectric constants,
for inorganic materials across 72 supported elements.
By leveraging transfer learning, 
the \textit{PFP} was treated as the frozen base model to 
encode material graphs with integrated elemental and structural information.
A light-weight trainable equivariant readout module is connected 
to integrate information in different tensorial orders for the final outputs.
It was discovered that embedded features from deeper GCN layers in \textit{PFP} 
contribute to better accuracy in predicting total dielectric constants.
After efficiently training the equivariant readout module,
our \textit{DTNet} model achieved superior results on the dielectric task in Matbench, and also higher accuracy in predicting the electronic,
ionic, total dielectric tensors compared with state-of-the-art models, PaiNN, M3GNet and MatTen, across various crystal system types.
To assess the model's capability in discovering novel dielectric materials,
we utilized it to identify high-score candidates near the tail of the training distribution for two distinct design targets by conducting virtual screening on total static dielectric constants.
For applications requiring both high dielectric constants and large band gaps in microelectronic manufacturing,
the model successfully identified the top three dielectric materials,
Cs\textsubscript{2}Ti(WO\textsubscript{4})\textsubscript{3} (mp-1226157, $E_g=2.83 \mathrm{eV}, \varepsilon=180.89$),
RbNbWO\textsubscript{6} (mp-1219587, $E_g=2.93 \mathrm{eV}, \varepsilon=155.57$) and
Ba\textsubscript{2}SmTaO\textsubscript{6} (mp-1214622, $E_g=3.36 \mathrm{eV}, \varepsilon=93.81$),
from a comprehensive pool of over 14,000 candidates, with 35 DFPT calculations performed.
For dielectrics with high anisotropy, the top three materials (including two 3D materials) identified by the model exhibit high anisotropic ratios of 128.890 (CsZrCuSe\textsubscript{3}, mp-7152), 96.763 (SeI\textsubscript{2}, mp-861871) and 61.026 (BaNiO\textsubscript{3}, mp-19138), respectively, validated by 41 DFPT calculations.
Promising candidates that significantly outperform the training distribution were successfully identified in both tasks.

Our work highlights that pretrained equivariant graph neural networks, which capture higher-order tensor representations, 
can incorporate common compositional and structural information, 
leading to improved accuracy in predicting properties of different orders. 
This approach also represents a viable alternative for enhancing prediction accuracy for other higher-order properties such as polarizability, multi-poles, and elasticity, 
thereby accelerating new material discoveries for specific applications.

\section{Methods}

\subsection{Data Source}

The Materials Project (v2023.11.1) contains a dataset of 7,277 dielectric tensors.
The dielectric properties are calculated using the Vienna Ab-Initio Simulation Package (VASP version 5.3.4),
employing the generalized gradient approximation GGA/PBE exchange-correlation functional~\cite{perdew1996generalized}
with the +U correction~\cite{dudarev1998electron,jain2011high} to account for electron-electron interactions within the transition metal orbitals.
Projector Augmented Wave pseudopotentials~\cite{blochl1994projector,kresse1999ultrasoft} were also included.
In the majority of cases, the dielectric constants obtained from these calculations exhibit a relative deviation of less than +/- 25\% when compared to experimental values at room temperature.
For each structure, it contains the value of the two components, namely, the electronic component and the ionic component.
The total dielectric tensor can be calculated by summing up the two components.
We clean up the dataset by filtering out dielectric constants that contain any element out of the range of $[-10, 100]$.
Structures that contain elements not supported by PFP (shown in Fig.~\ref{fig:data_dist}(c)) are cleaned up, 
and finally there are 6,648 structures retained for model training.

The dataset is randomly split into the training, validation and test data in the ratio of 80\%, 10\% and 10\%, respectively.
For each experiment conducted in the present study, we do 5 different random splits first and we report their average performance as well as the deviation of the 5 runs to reduce the impact of random.

\subsection{Model Architectures}

Once the model is trained on large potential data, we extract the intermediate knowledge $\boldsymbol{a}_s^n$, $\boldsymbol{a}_v^n$, $\boldsymbol{a}_t^n$, $\boldsymbol{b}_s^n$, and $\boldsymbol{b}_v^n$, 
from a pretrained \textit{PFP}-$\mathrm{L}_n$ as the input for our equivariant readout layers.

\textbf{Overview}\quad The input materials are preprocessed
to identify appropriate neighbor atoms under the periodic boundary conditions for graph construction.
Each material is represented by a graph $\mathcal{G}=(\mathcal{V},\mathcal{E})$.
The initialized atom attributes contain three types of information with different ranks:
$\mathcal{V}=\{(\boldsymbol{a}_s^0, \boldsymbol{a}_v^0, \boldsymbol{a}_t^0)_i\}_{i=1:N^a}$,
where $\boldsymbol{a}_s^0\in\mathbb{R}^{C_s^a}$, $\boldsymbol{a}_v^0\in\mathbb{R}^{C_v^a\times 3}$, $\boldsymbol{a}_t^0\in\mathbb{R}^{C_t^a\times 3\times 3}$,
and $N^a$ is the number of atoms.
$\boldsymbol{a}_s^0$ is initialized by looking up a predefined table, 
in which each element is mapped to a high-dimension encoding with summation equal to the atomic number divided by 2.
Both atom vector $\boldsymbol{a}_v^0$ and atom tensor $\boldsymbol{a}_t^0$ are initialized to be zeros.
Two types of bond attributes are prepared:
$\mathcal{E}=\{(\boldsymbol{b}_s^0, \boldsymbol{b}_v^0)_k\}_{k=1:N^b}$, 
in which $\boldsymbol{b}_s^0\in\mathbb{R}^{C_s^b}$, $\boldsymbol{b}_v^0\in\mathbb{R}^{C_v^b\times 3}$,
$N^b$ is the number of bonds between atom pairs within the cutoff radius $R_c$,
and $C_s^a$, $C_v^a$, $C_t^a$, $C_s^b$, $C_v^b$ are the corresponding number of channels for each feature.
Both $\boldsymbol{b}_v^0$ and $\boldsymbol{b}_s^0$ are filled with zeros initially in this work.

\textit{TeaNet}~\cite{takamoto2022teanet} is used as the base GNN architecture in \textit{PFP}~\cite{takamoto2022towards, takamoto2023towards}.
The \textit{TeaNet} architecture incorporates a second-order Euclidean tensor
for high-order geometry interaction and performs equivariant graph convolutions using its information.
The original material~\cite{takamoto2022teanet} provides a step-by-step explanation for \textit{TeaNet} implementation for more details.
\textit{PFP} has several modifications on \textit{TeaNet}, such as introducing the Morse-style two-body potential term and so on, see the corresponding material~\cite{takamoto2022towards} for details.
\textit{PFP} holds the scalar features $\boldsymbol{a}_s$, $\boldsymbol{b}_s$ invariant and the vector/tensor features $\boldsymbol{a}_v$, $\boldsymbol{a}_t$, $\boldsymbol{b}_v$ equivariant to rotations/inversions.
\textit{PFP} calculates local interactions in each of its GCN layers.
5-layer GCN layers are constructed with the cutoff radius $R_c$ as 3, 3, 4, 6 and 6\AA, respectively.
In other words, latent embeddings in the earlier convolutional layers hold less distant information.
As a result, the maximum distance of atomic interactions counted in \textit{PFP} is 22{\AA}
after the 5-layer information propagation. 
The intermediate atom/bond features $\boldsymbol{a}_s^n$, $\boldsymbol{a}_v^n$, $\boldsymbol{a}_t^n$, $\boldsymbol{b}_s^n$, and $\boldsymbol{b}_v^n$ extracted from the $n$-th GCN layer are used as the input to the designed \textit{DTNet} which contains $M$ equivariant blocks and a readout neural network.

\textbf{Neural network definition}\quad We denote the one-layer linear transformation without the bias term as
\begin{equation}
	\mathcal{L}^k: x\mapsto\boldsymbol{W_k}x
\end{equation}
and one layer of the perceptron model as
\begin{equation}
	\mathcal{L}_g^k: x\mapsto g(\boldsymbol{W_k}x+\boldsymbol{b_k})
\end{equation}
where $\boldsymbol{W_k}$ and $\boldsymbol{b_k}$ are learnable parameters, $g(x)$ is the activation function which can be substituted by $s(x)$ as the SiLU activation function and $\sigma(x)$ as the sigmoid activation function.
The SiLU activation function $s(x)$~\cite{hendrycks2016gaussian,elfwing2018sigmoid,ramachandran2017swish} is defined by the sigmoid function $\sigma(x)$ multiplied by its input $x$
\begin{equation}
	s(x) = x * \sigma(x) = \frac{x}{1+e^{-x}}
\end{equation}
Thus the $K$-layer MLP with $s(x)$ in the intermediate layers and without the activation function in the final layer can be expressed as
\begin{equation}\label{eq:mlp}
	\xi^K(x) = (\mathcal{L}^K(x)\circ\mathcal{L}_s^{K-1}\circ...\mathcal{L}_s^1)(x)
\end{equation}
And the $K$-layer gated MLP~\cite{liu2021pay} is represented by
\begin{equation}
	\phi^K(x) = ((\mathcal{L}^K(x)\circ\mathcal{L}_s^{K-1}\circ...\mathcal{L}_s^1)(x))\odot((\mathcal{L}_\sigma^K(x)\circ\mathcal{L}_s^{K-1}\circ...\mathcal{L}_s^1)(x))
\end{equation}
where $\odot$ denotes the element-wise product.
It comprises two networks, i.e., a normal MLP defined as defined in Eq.\ref{eq:mlp} and a gate network with the final layer activated by a sigmoid function.

\textbf{Equivariant block}\quad The equivariant block takes $\boldsymbol{a}_s^{(n,m)}$, $\boldsymbol{a}_v^{(n,m)}$, $\boldsymbol{a}_t^{(n,m)}$, $\boldsymbol{b}_s^n$, and $\boldsymbol{b}_v^n$ as inputs, 
where $n$ denotes the GCN layer number in \textit{PFP}, and $m$ denotes the layer number of the equivariant block.
The perceptron models are applied to the atom scalar features $\boldsymbol{a}_s^{(n,m)}$ and edge scalar features $\boldsymbol{b}_s^n$, respectively,
\begin{equation}
	\boldsymbol{a}_{s1}^{(n,m)} = \xi_{a_s1}^K(\boldsymbol{a}_s^{(n,m)})
\end{equation}
\begin{equation}
	\boldsymbol{b}_{s1}^{(n,m)} = \xi_{b_s1}^K(\boldsymbol{b}_s^{(n,m)})
	\label{eq: edge scalar}
\end{equation}
To keep the equivariance of vector features and tensor features, the linear transformations without the bias term are applied to the atom vector features $\boldsymbol{a}_v^{(n,m)}$, atom tensor features $\boldsymbol{a}_t^{(n,m)}$, and bond vector features $\boldsymbol{b}_v^n$.
\begin{equation}
	\boldsymbol{a}_{v1}^{(n,m)} = \mathcal{L}_{a_v1}(\boldsymbol{a}_v^{(n,m)})
\end{equation}
\begin{equation}
	\boldsymbol{a}_{t1}^{(n,m)} = \mathcal{L}_{a_t1}(\boldsymbol{a}_t^{(n,m)})
\end{equation}
\begin{equation}
	\boldsymbol{b}_{v1}^n = \mathcal{L}_{b_v1}(\boldsymbol{b}_v^n)
	\label{eq: edge vector}
\end{equation}
The neighbor set of atom $i$ is denoted as $\mathcal{N}_i$. 
Let $\boldsymbol{a}_{s1,i}^{(n,m)}$ denote the scalar feature of atom $i$,
and $\boldsymbol{b}_{s1,\{i,j\}}^n$ denote the scalar feature of the bond connecting atom $i$ and atom $j$. 
Other vector and tensor features are represented in a similar manner.
The bond scalar features $\boldsymbol{b}_{s1,\{i,j\}}^n$ and vector features $\boldsymbol{b}_{v1,\{i,j\}}^n$ are aggregated to atom scalar features $\boldsymbol{a}_{s1,i}^{(n,m)}$ and vector features $\boldsymbol{a}_{v1,i}^{(n,m)}$ accordingly, thus for each atom $i$
\begin{equation}
	\boldsymbol{a}_{s2,i}^{(n,m)} = \boldsymbol{a}_{s1,i}^{(n,m)} + \sum_{j\in\mathcal{N}_i}\boldsymbol{b}_{s1,\{i,j\}}^n
\end{equation}
\begin{equation}
	\boldsymbol{a}_{v2,i}^{(n,m)} = \boldsymbol{a}_{v1,i}^{(n,m)} + \sum_{j\in\mathcal{N}_i}\boldsymbol{b}_{v1,\{i,j\}}^n
\end{equation}
Then the Frobenius norm $\Vert\cdot\Vert$ of node vector features and tensor features are combined with the node scalar features and fed into another perceptron model.
\begin{equation}
	\boldsymbol{a}_{s3}^{(n,m)} = \xi_{a_s3}^K(\mathrm{concat}(\boldsymbol{a}_{s2}^{(n,m)} + \Vert\boldsymbol{a}_{v2}^{(n,m)}\Vert+\Vert\boldsymbol{a}_{t2}^{(n,m)}\Vert)
\end{equation}
Inspired by the gate nonlinearity design~\cite{weiler20183d}, 
the interaction output of different-rank node features are split, and the vector and tensor components are multiplied by the sigmoid activation function to form a gate equivariant nonlinearity for the transformed input of the vector feature and tensor feature, respectively.
\begin{equation}
	\boldsymbol{a}_{s4}^{(n,m+1)}, \boldsymbol{a}_{v3}^{(n,m)}, \boldsymbol{a}_{v3}^{(n,m)} = \mathrm{split}(\boldsymbol{a}_{s3}^{(n,m)})
\end{equation}
\begin{equation}
	\boldsymbol{a}_{v4}^{(n,m+1)} = \sigma(\boldsymbol{a}_{v3}^{(n,m)})\odot\mathcal{L}_{a_v2}(\boldsymbol{a}_v^{(n,m)})
\end{equation}
\begin{equation}
	\boldsymbol{a}_{t4}^{(n,m+1)} = \sigma(\boldsymbol{a}_{t3}^{(n,m)})\odot\mathcal{L}_{a_t2}(\boldsymbol{a}_t^{(n,m)})
\end{equation}
Finally, the updated $\boldsymbol{a}_{s4}^{(n,m+1)}$, $\boldsymbol{a}_{v4}^{(n,m+1)}$ and $\boldsymbol{a}_{t4}^{(n,m+1)}$ are used as the input for the next equivariant readout block.

\textbf{Readout}\quad The final layer output, i.e., layer $M$, of the equivariant blocks $\boldsymbol{a}_s^{(n,M)}$, $\boldsymbol{a}_v^{(n,M)}$ and $\boldsymbol{a}_t^{(n,M)}$ are aggregated for the tensorial property prediction.
Since the dielectric constant is an intensive property, we take the average aggregation of the atom features in the graph representation of materials.
The gated MLP applied to the scalar features $\boldsymbol{a}_s^{(n,M)}$ is found to help improving the accuracy compared with the normal MLP.
Therefore, the atom scalar features is processed by
\begin{equation}
	\boldsymbol{a}_s' = \phi^K(\boldsymbol{a}_s^{(n,M)})
\end{equation}
The channels of atom vector and tensor features are compressed to 1 by
\begin{equation}
	\boldsymbol{a}_v' = \mathcal{L}_{a_v3}(\boldsymbol{a}_v^{(n,M)})
\end{equation}
\begin{equation}
	\boldsymbol{a}_t' = \mathcal{L}_{a_t3}(\boldsymbol{a}_t^{(n,M)})
\end{equation}
Finally, we construct the dielectric tensors using
\begin{equation}
	\boldsymbol{\varepsilon} = \frac{1}{N^a}\sum^{N^a}(\boldsymbol{a}_s'\boldsymbol{I}_3 + \boldsymbol{a}_v' \otimes \boldsymbol{a}_v' + \boldsymbol{a}_t')
\end{equation}
where $\boldsymbol{I}_3$ is the $3\times3$ identity tensor, and $\otimes$ represents the outer product operation.
The first models isotropic, and the final two terms add the anisotropic components of the dielectric tensor.
Therefore, the final output of our model is the $3\times3$ dielectric tensor following the equivariance of the materials input.

\subsection{Hyperparameters}
The \textit{DTNet} model contains 2 equivariant blocks (Fig. \ref{fig:equi}(a)).
All the latent dimensions in MLP in Fig. \ref{fig:equi}(b) and (c) are set to 128 with a dropout rate of 0.2.
The batch size is set to 64 during training. 
The AdamW optimizer~\cite{loshchilov2017decoupled} is utilized to optimize parameters with $\beta_1=0.9, \beta_2=0.999, \lambda=0.01$.
The learning rate starts at $1\times10^{-4}$ and a cosine annealing schedule is applied for tighter convergence.
The gradient norm is clipped at 2.0 to stabilize the training.
Early stopping is applied during training if no improvement on the validation accuracy is observed after 200 epochs.

\subsection{Virtual Screening}

We performed virtual screening based on iterative active learning to search for candidates for the two problems.
The acquisition function uses an upper confidence bound (UCB) to balance exploration and exploitation during multi-round screening, defined by
\begin{equation}
    a(\chi; \kappa) = \mu(\chi) + \kappa\cdot\sigma(\chi)
    \label{eq:active}
\end{equation}
where $\chi$ is the input crystal structure, $\kappa$ is a hyperparameter controlling the trade-off and here $\kappa=1.5$ is selected for both tasks.
$\mu(\chi)$ and $\sigma(\chi)$ represent the mean and standard deviation of polycrystalline dielectric constants and anisotropy ratios in the two respective tasks.
In each task, we select the top 20 highest $a(x;\kappa=1.5)$ candidates with unit cell sizes of up to 20 atoms for validation by DFPT in each round in consideration of computational budget constraints.
After each round, the dataset was updated with the newly validated results, and the next iteration was performed.
In this case, finally, there were 60 new structures validated by DFPT computations across 3-round active learning for each discovery task.

\subsection{DFPT Configuration}
The density functional perturbation theory (DFPT) validation of the screened results was performed using the Vienna Ab initio Simulation Package (VASP)~\cite{kresse1996efficient}.
For the calculations, the generalized-gradient approximation (GGA) with the Perdew-Burke-Ernzerhof (PBE) functional was employed, augmented by Hubbard U corrections with the same settings on the MP public documentation~\cite{materialsproject}.
The projector augmented wave (PAW) method~\cite{kresse1999ultrasoft} was used to represent the core and valence electrons.
The plane-wave basis set was truncated at an energy cutoff of 600 eV.
The k-point density of 3000 k-points per reciprocal atom was used.
A Gaussian smearing with a width of 0.05 eV was applied to account for partial occupancies in the electronic structure.
A tight global electronic relaxation convergence criterion is applied with an energy difference tolerance of $1\times10^{-7}$ eV.
The electronic minimization utilizes a fairly robust mixture of the blocked-Davidson and RMM-DIIS algorithms to achieve efficient convergence.
The precision setting was configured to "Accurate" to enhance the overall precision of the calculations.

\section*{Acknowledgments}
The authors thank So Takamoto for fruitful discussions and helpful comments.
The authors thank Chikashi Shinagawa for assistance in DFPT computations using VASP.
The research is supported by Preferred Networks Inc.

\section*{Competing interests}

The authors declare the following financial interests/personal relationships which may be considered as potential competing interests: 
Zetian Mao was an intern/part-time employee at Preferred Networks, Inc. during this study.
Jethro Tan and Wenwen Li are full-time employees working at Preferred Networks, Inc. 

\section*{Data availibility}

The materials with dielectric constants are freely accessible using The Materials Project API (\url{https://next-gen.materialsproject.org/api}).
Our preprocessed data from MP used in this study and our validated DFPT results are provided at \url{https://github.com/pfnet-research/dielectric-pred}.

\section*{Code availability}

The implementation code of \textit{DTNet} is available at \url{https://github.com/pfnet-research/dielectric-pred}. The pretrained model \textit{PFP} is provided in the proprietary software named Matlantis. The code and trained parameters are not open-source. The related information of latest-version \textit{PFP} can be found at \url{https://matlantis.com/news/pfp-validation-for-public-v5-0-0}.

\section*{Author contribution}
Z. M. implemented the codes and conducted experiments.
W. L, J. T. and Z. M. conceived the idea.
W. L. and J. T. provided technical support of \textit{PFP}, and supervised the research.
All authors wrote and confirmed the manuscript.

\bibliographystyle{unsrt}  
\bibliography{references}

\end{document}